\documentclass{ckm}                 

\usepackage{txfonts}            

\confname{
  Workshop on the CKM Unitarity Triangle, IPPP Durham, April 2003
}

\title{
  Status and Prospects for Measurements of $\phi_3$ from $B \to D X$ Decays
}

\author{
  T. Gershon\thanks{
    Supported by the Japan Society for the Promotion of Science.
  }
}
\address
{High Energy Accelerator Research Organization (KEK),
  1-1 Oho, Tsukuba, Ibaraki, 305-0801, Japan.}

\begin{document}

\begin{abstract}
Methods for extracting the Unitarity Triangle angle $\phi_3$ 
from $B$ decays to final states incluing a $D$ meson are reviewed.
The current experimental status and future prospects are summarized.
\end{abstract}

\maketitle

\setcounter{footnote}{0}

\section{Introduction}
The angle of the Unitarity Triangle $\phi_3$ is defined as 
$\phi_3 \equiv \mathrm{arg} 
\left( - V_{ud}V_{ub}^* / V_{cd}V_{cb}^* \right)$.\footnote{
  The translation to the other familiar notation is:
  $\phi_1 \equiv \beta$, $\phi_2 \equiv \alpha$, $\phi_3 \equiv \gamma$.
}
Therefore, in order to observe its effect, interference between
$b \to u$ and $b \to c$ processes is required.
Decays of the type $B \to D X$, 
where $X$ represents one or more light mesons,
which can be mediated by both these transitions
are thus a natural environment in which to attempt to measure $\phi_3$.
These modes are typically theoretically clean,
as only tree diagrams are involved; 
furthermore, they tend to have reasonably large branching fractions.
However, the modes with the largest decay rates also tend to have 
the $b \to u$ amplitude highly suppressed relative to the
$b \to c$ transition, 
making observation of the interference effects difficult.
In addition, the experimental efficiency to reconstruct the $D$ meson
reduces the yield of events.
The challenge, both experimental and theoretical, 
is to find approaches which obviate these obstacles.

In what follows, the magnitude of the ratio of the 
$b \to u$ (suppressed) to $b \to c$ (favoured)
amplitudes will be denoted as $R$, 
whilst the relative strong phase will be denoted as $\delta$.
Averaging over charge conjugate states is implied for
branching fractions (denoted by ${\mathcal B}$);
in other places the meaning should be clear from the context.

\section{\boldmath{$\phi_3$} From \boldmath{$B \to D X_s$}}

\subsection{Phenomenology}

\begin{figure}
  \hbox to\hsize{
    \hss
    \includegraphics[width=\hsize]{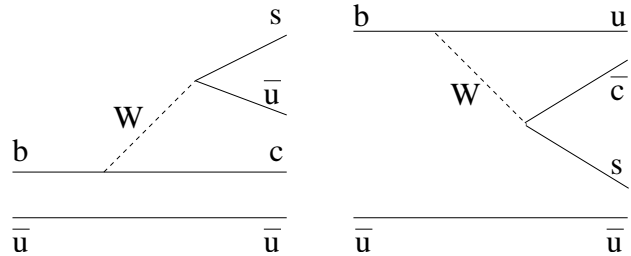}
    \hss
  }
  \caption{
    Diagrams for (left) $B^- \to D^0 K^-$, 
    (right) $B^- \to \bar{D}^0 K^-$.
  }
  \label{fig:feynman}
\end{figure}

\begin{figure}
  \hbox to\hsize{
    \hss
    \includegraphics[width=\hsize]{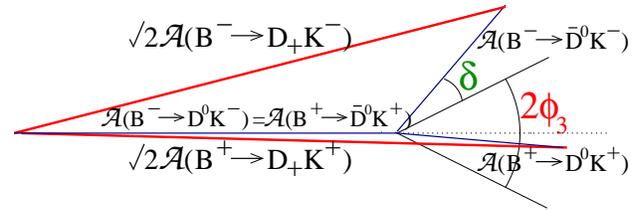}
    \hss
  }
  \caption{Amplitude relations in $B \to D_{CP}K$ decays.}
  \label{fig:glw_triangle}
\end{figure}

Of methods of this type, one of the first to appear in the literature
was proposed by Gronau, London and Wyler (GLW) \cite{glw}.
Most methods to extract $\phi_3$ from $B \to D X_s$ decays
can be considered as variants of this technique.
Diagrams 
for the $b \to c$ transition $B^- \to D^0 K^-$
and the $b \to u$ transition $B^- \to \bar{D}^0 K^-$
are shown in Fig.~\ref{fig:feynman}.
If the $D$ meson is reconstructed in a state to which
both $D^0$ and $\bar{D}^0$ can decay, 
the diagrams will interfere, resulting in $CP$ violating observables.
In the GLW method, the $D$ meson is reconstructed in a $CP$ eigenstate.
By defining the neutral $D$ $CP$ eigenstates
$D_{\pm} = \frac{1}{\sqrt{2}} \left( D^0 \pm \bar{D}^0 \right)$,
the amplitude relations can be drawn as shown in Fig.~\ref{fig:glw_triangle}.
Simple trigonometry can then be used to derive the $CP$ asymmetries,
\begin{eqnarray}
A_{D_+K^-} & = & 
\frac{2 R_{DK^-} \sin(\delta_{DK^-})\sin(\phi_3)}
{1+R_{DK^-}^2+2R_{DK^-}\cos(\delta_{DK^-})\cos(\phi_3)} \\
A_{D_-K^-} & = & 
\frac{-2 R_{DK^-} \sin(\delta_{DK^-})\sin(\phi_3)}
{1+R_{DK^-}^2-2R_{DK^-}\cos(\delta_{DK^-})\cos(\phi_3)}.
\end{eqnarray}

It is apparent that in order for $CP$ violation to be observable,
$\delta_{DK^-}$ (in addition to $\phi_3$) must be non-zero.
Furthermore, in order to extract the value of $\phi_3$,
either or both of $R_{DK^-}$ and $\delta_{DK^-}$ must be known.
One might hope to measure $R_{DK^-}$
from the ratio of branching fractions,
${\mathcal B}\left( B^- \to \bar{D}^0 K^- \right) / {\mathcal B}\left( B^- \to D^0 K^- \right)$,
where the $D^0$/$\bar{D}^0$ are reconstructed in flavour specific final states.
Ideally, semi-leptonic $D$ decays could be used for such a measurement.
Unfortunately, enormous backgrounds would have to be overcome
in such an analysis, which to date have made this approach unfeasible.
Alternatively, hadronic decays of the type $D^0 \to K^- \pi^+$ could be
used to tag the flavour of the $D$.  
However, doubly Cabibbo-suppressed decays of the type $D^0 \to K^+ \pi^-$
have to be taken into account.  
Since the products of amplitudes
$
{\mathcal A}\left( B^- \to \bar{D}^0 K^- \right) 
\times 
{\mathcal A}\left( \bar{D}^0 \to K^+ \pi^- \right)
$
and
$
{\mathcal A}\left( B^- \to D^0 K^- \right) 
\times 
{\mathcal A}\left( D^0 \to K^+ \pi^- \right)
$
are predicted to be similar in magnitude,
the doubly Cabibbo-suppressed $D$ decays preclude this approach.

Nonetheless, there is sufficient information to extract $\phi_3$,
if the ratios of branching fractions to $CP$ eigenstates
and to quasi-flavour specific (favoured) states are included~\cite{gronau98}.
It is convenient to normalize each of these branching fractions
to the $B^- \to D^0 \pi^-$ rates 
(highly suppressed contributions from $B^- \to \bar{D}^0 \pi^-$ 
can be neglected).
In this way some systematic effects can be removed,
and additionally the branching fraction ratios
${\mathcal B}\left(B \to D K\right)/{\mathcal B}\left(B \to D \pi\right)$
can be used to test the factorization hypothesis.
Defining the double ratios ${\mathcal R}_{\pm}$ as
\begin{equation}
{\mathcal R}_{D_{\pm}K^-} = 
\frac{
  {\mathcal B}\left( B^- \to D_{\pm} K^-   \right)/
  {\mathcal B}\left( B^- \to D_{\pm} \pi^- \right)
}{
  {\mathcal B}\left( B^- \to D^0 K^-   \right)/
  {\mathcal B}\left( B^- \to D^0 \pi^- \right)
},
\end{equation}
it is again a matter of mere trigonometry to derive
\begin{equation}
{\mathcal R}_{D_{\pm}K^-} =  
1 + R_{DK^-}^2 \pm 2 R_{DK^-} \cos(\delta_{DK^-}) \cos(\phi_3).
\end{equation}

Since $A_{D_+K^-}{\mathcal R}_+ = -A_{D_-K^-}{\mathcal R}_-$,
it can be seen that there are three independent measurements
and three unknowns $(R_{DK^-}, \delta_{DK^-}, \phi_3)$
and hence there is enough information to extract $\phi_3$.
An eight-fold discrete ambiguity remains, however~\cite{ambiguity}.

A remaining problem with this technique is that the size 
of the $CP$ violating observable depends on the size of $R_{DK^-}$.
As the $b \to c$ transition is colour-allowed whilst
the $b \to u$ transition is colour-suppressed,
early predictions for this value were ${\mathcal O}\left(0.1\right)$.
Since the observation of larger than expected colour-suppressed
decay amplitudes~\cite{col_supp}, these predictions have been revised
upwards, and an optimistic estimate is now 
${\mathcal O}\left(0.2\right)$~\cite{gronau03}.

Note that the formalism above has neglected possible effects
from $CP$ violation and mixing in the $D$ sector.
A more thorough treatment can take such effects into account~\cite{dmixing}.

\subsection{\boldmath{$B^- \to D_{CP}K^-$} Experimental Status}

\begin{figure}
  \hbox to\hsize{
    \hss
    \includegraphics[width=\hsize]{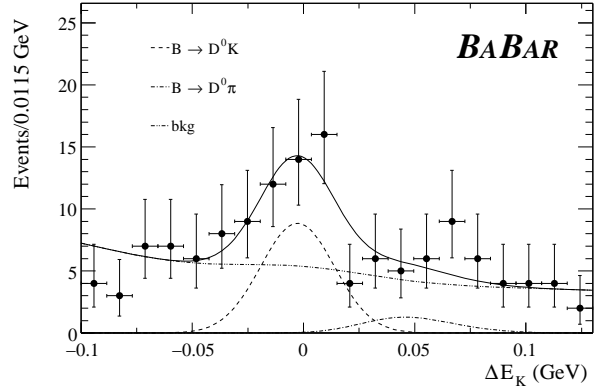}
    \hss}
  \caption{
    Yield of $B^{\mp} \to D_+ K^{\mp}$ from BaBar.
    A contribution from $B^{\mp} \to D_+ \pi^{\mp}$ can be seen at positive
    $\Delta E_K$ (the kaon mass is assumed for the primary track).
  }
  \label{fig:babar_dcpk}
\end{figure}

\begin{figure}
  \hbox to\hsize{
    \hss
    \includegraphics[width=\hsize]{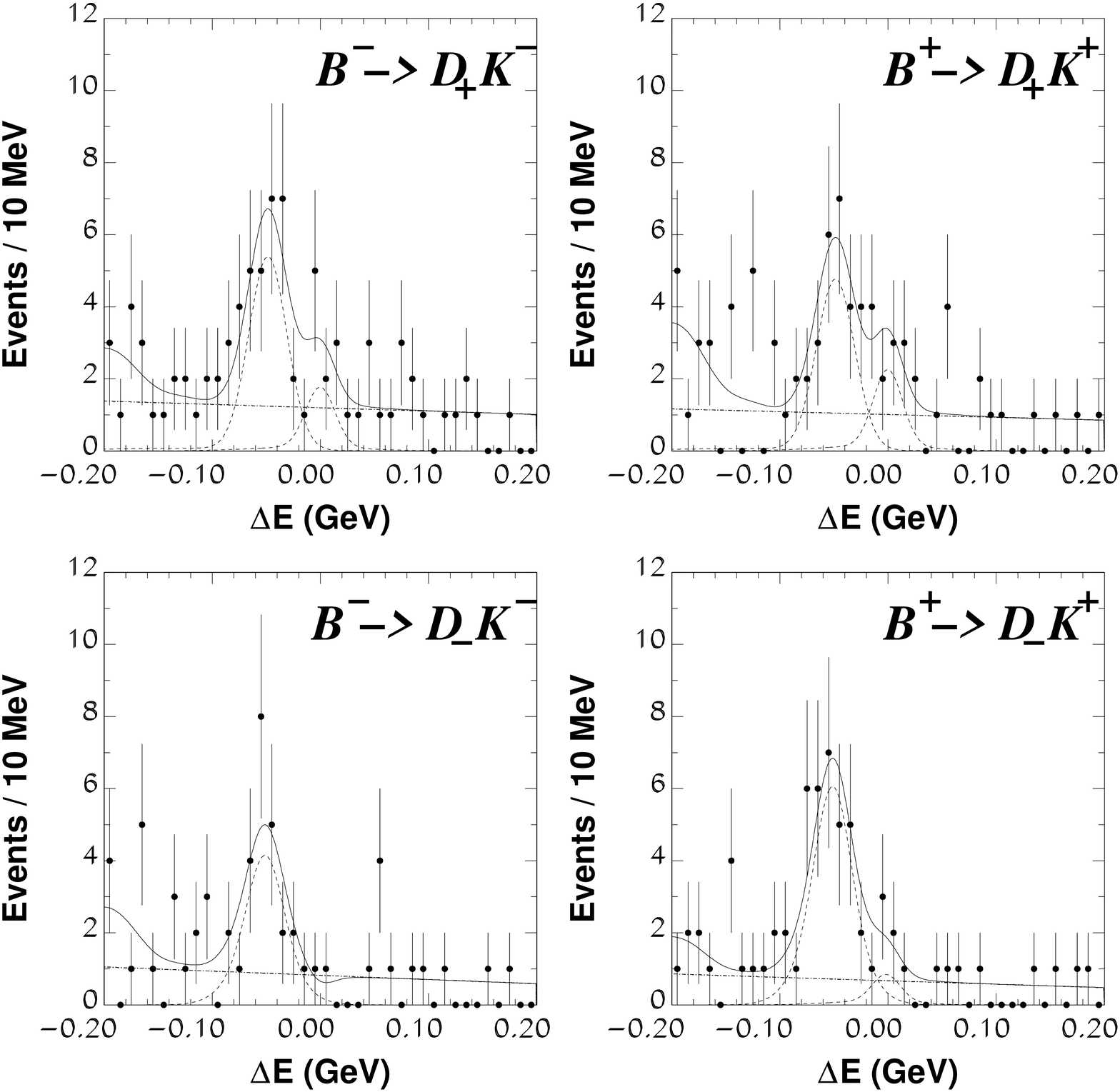}
    \hss}
  \caption{
    Yields of $B^{\mp} \to D_{\pm} K^{\mp}$ from Belle.
    Contributions from $B^{\mp} \to D_{\pm} \pi^{\mp}$ 
    can be seen around $\Delta E = 0$
    (the pion mass is assumed for the primary track).
  }
  \label{fig:belle_dcpk}
\end{figure}

BaBar and Belle have released results on $B^- \to D_{CP} K^-$ decays.
As shown in Fig.~\ref{fig:babar_dcpk},
BaBar reconstruct the $D$ meson in the $CP = +1$ eigenstate
$K^+K^-$ with $75\ {\rm fb}^{-1}$ of data~\cite{babar_dkkk} 
(recent results using $D_+ \to \pi^+\pi^-$~\cite{babar_dpipik}
are not included here).
Belle use $78\ {\rm fb}^{-1}$ of data~\cite{belle_dcpk},
reconstruct $K^+K^-$ and $\pi^+\pi^-$ for the $CP$-even decays,
and in addition reconstruct the $D$ in the $CP$-odd final states
$K_S\pi^0$, $K_S\phi$, $K_S\omega$, $K_S\eta$ and $K_S\eta^{\prime}$.
These are shown in Fig.~\ref{fig:belle_dcpk}.
The results are summarized in Table~\ref{tab:dcpk}.
\begin{table}
  \hbox to\hsize{
    \hss
  \begin{tabular}{ccc}
    \hline
    & BaBar & Belle \\
    \hline
    ${\mathcal R}_+$ & $\frac{7.4 \pm 1.7 \pm 0.6}{8.31 \pm 0.35 \pm 0.20}$ 
                     & $ 1.25 \pm 0.25 \pm 0.14$ \\
    ${\mathcal R}_-$ & - 
                     & $ 1.41 \pm 0.27 \pm 0.15$ \\
    $A_{D_+K^-}$     & $0.17 \pm 0.23^{+0.09}_{-0.07}$ 
                     & $+0.06 \pm 0.19 \pm 0.04$ \\
    $A_{D_-K^-}$     & - 
                     & $-0.18 \pm 0.17 \pm 0.05$ \\
    \hline
  \end{tabular}
    \hss
  }
  \caption{
    Summary of $B^- \to D_{CP}K^-$ results.
    BaBar~\cite{babar_dkkk} give results only for single ratios;
    no attempt is made here to estimate the correlation of systematic errors.
  }
  \label{tab:dcpk}
\end{table}
No significant $CP$ asymmetry has yet been observed in these modes.
It is clear that substantially larger data sets will be required 
in order to measure $\phi_3$.

Note that a number of similar decay modes, 
which can be denoted generically as $B^- \to D^{(*)}_{CP}K^{(*)-}$,
can be used for essentially the same analysis.
However, these tend to have smaller reconstruction efficiencies,
and so smaller yields are obtained, 
increasing the statistical error in these modes.

\subsection{\boldmath{$B^- \to D K \pi$}}

A recent extension to the above method involves considering
the entire Dalitz plot of $B^- \to D_{CP} K^- \pi^0$ decays~\cite{petersen}.
Non-resonant contributions can be produced by colour-allowed
transitions of both $b \to c$ and $b \to u$.
Therefore, the ratio of amplitudes $R_{D K \pi}$ is expected to be large,
resulting in augmented interference effects.
However, the $B^- \to D K \pi$ Dalitz plot is likely to be dominated
by resonances, and the non-resonant contribution may be rather small.
The expected resonant structures include 
$B^- \to D^{*}_{CP} K^-$, $B^- \to D_{CP} K^{*-}$ and 
$B^- \to D_s^{**-} \pi^0$.
The first two have the $b \to u$ transition colour-suppressed,
exactly as before; the latter is a pure $b \to u$ transition.
Hence, if the Dalitz plot is dominated by these resonances,
it appears that there is no large improvement over the quasi-two body analysis,
although interference between $D^{*}_{CP} K^-$ and $D_{CP} K^{*-}$
may help to resolve ambiguities.
If, however, there is a large non-resonant component,
or at least that there is some reasonably well-populated region 
of the Dalitz plot with a large value of $R_{D_{CP}K^-\pi^0}$,
this method will allow extraction of $\phi_3$ with only a single ambiguity.
Data analysis will reveal whether this condition is satisfied.

\subsection{The ADS Method}

One variation of the GLW method, 
proposed  by Atwood, Dunietz and Soni (ADS)~\cite{ads},
makes use of the doubly Cabibbo-suppressed $D$ decays
which prevented the measurement of $R_{DK^-}$.
As previously noted, the contributions to
$B^- \to (K^+\pi^-)_{D} K^-$ from
$B^- \to D^0 K^-$ followed by $D^0 \to K^+ \pi^-$, and from
$B^- \to \bar{D}^0 K^-$ followed by $\bar{D}^0 \to K^+ \pi^-$
are expected to be similar in size, and therefore the
$CP$ asymmetry may be ${\mathcal O}\left(1\right)$.
Whilst additional information will be needed to extract the value of $\phi_3$,
a measurement of non-zero $CP$ asymmetry in such a mode would
be a clear signal of direct $CP$ violation.
Unfortunately, these modes are rather rare, 
and to date none have been observed.
However, it may be possible to increase statistics 
by using inclusive $D$ decays
of the form $B^- \to (K^+ X)_{D} K^-$~\cite{atwood}.

\subsection{\boldmath{$\bar{B}^0 \to D \bar{K}^{(*)0}$}}

Another alternative can be found by considering neutral $B$ decays.
The amplitudes 
$\bar{B}^0 \to D^0 \bar{K}^{(*)0}$ ($b \to c$) and
$\bar{B}^0 \to \bar{D}^0 \bar{K}^{(*)0}$ ($b \to u$) 
are both colour-suppressed,
leading to a value of $R_{D\bar{K}^{(*)0}}$ as large as 
$0.4$~\cite{kayserlondon}.
In the case that $\bar{K}^{*0} \to K^-\pi^+$,
the flavour of the kaon is tagged by its decay products.
Therefore, precise measurements of the rates and $CP$ asymmetries for
$\bar{B}^0 \to D_{\pm} \bar{K}^{*0}$ and of the rates for
$\bar{B}^0 \to D^0 \bar{K}^{*0}$ and 
$\bar{B}^0 \to \bar{D}^0 \bar{K}^{*0}$
will allow extraction of $\phi_3$~\cite{dunietz91}.
When the final state includes $K_S \to \pi^+ \pi^-$
the contributions from $B^0$ and $\bar{B}^0$ decays cannot be disentangled
without tagging the flavour of the decaying $B$.
In $e^+e^-$ $B$ factories, this is achieved by
identifying the flavour of the other $B$ in the $\Upsilon(4S) \to B\bar{B}$
decay~\cite{tagging}, so dilution due to $B-\bar{B}$ mixing 
has to be taken into account.
An intriguing prospect in this case, 
is to study the time-evolution of the system~\cite{kayserlondon,fleischer03}.
Note, however, that all the final state particles originate 
from decays of secondary particles with have non-negligible lifetimes 
($D^0$, $K_S$),
complicating the determination of the $B$ vertex position.

\begin{figure}
  \hbox to\hsize{
    \hss
    \includegraphics[width=\hsize]{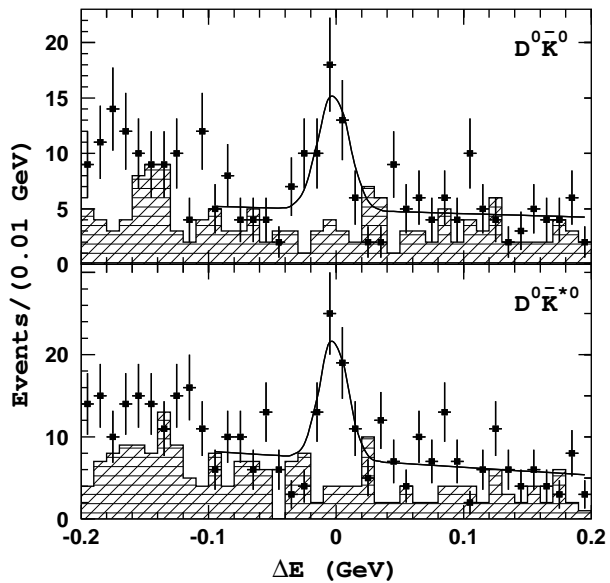}
    \hss}
  \caption{
    Observations of $\bar{B}^0 \to D^0 \bar{K}^{(*)0}$ by Belle.
    The hatched histograms show the distribution
    of events in a $D^0$ sideband region.
  }
  \label{fig:d0k0}
\end{figure}

As shown in Fig.~\ref{fig:d0k0}, 
Belle have recently observed the decays 
$\bar{B}^0 \to D^0 \bar{K}^0$ and $\bar{B}^0 \to D^0 \bar{K}^{*0}$
with branching fractions of 
$\left( 5.0^{+1.3}_{-1.2} \pm 0.6 \right) \times 10^{-5}$ and
$\left( 4.8^{+1.1}_{-1.0} \pm 0.5 \right) \times 10^{-5}$, 
respectively\cite{belle_d0k0}, using 
a data sample of $78\ {\rm fb}^{-1}$.
Significantly larger data samples will be required for analyses 
in which the $D^0$ is reconstructed in a $CP$ eigenstate.
However, if the ratio $R$ is large, as expected, 
the observation of $\bar{B}^0 \to \bar{D}^0 \bar{K}^{*0}$
should soon be within the reach of the $B$-factories.

\subsection{Multibody \boldmath{$D$} Decays}

Recently, another possible method to extract $\phi_3$ from the 
interference between $B^- \to D^0 K^-$ and $B^- \to \bar{D}^0 K^-$
has appeared in the literature~\cite{zupan}.
Here the $D$ is reconstructed in a multibody final state
which can be reached via both $D^0$ and $\bar{D}^0$ decays;
a typical example is $K_S \pi^+ \pi^-$.
At each point in the phase space (Dalitz plot for $K_S \pi^+ \pi^-$)
with contributions from both $D^0$ and $\bar{D}^0$,
the interference pattern will be different for $B^-$ and $B^+$ decays,
due to the weak phase $\phi_3$.
The phase space can be analysed in either a model independent manner,
in which case additional information will be required from 
charm factories ($\psi(3770) \to D\bar{D}$), 
or with some model dependence introduced by using Breit-Wigner forms 
for the resonances which contribute to the final state of interest.
In the model dependent method, 
additional information about the variation of the strong phase results
in only a single ambiguity.

Note that the structure of such multibody $D$ decays can be studied at 
$B$ factories using the large samples of $D$ mesons
which are tagged by $D^{*+} \to D^0 \pi^+$ decays.
Indeed, CLEO has performed a Dalitz plot analysis 
of tagged $D$ mesons decaying to the $K_S \pi^+ \pi^-$
final state~\cite{cleo_kspipi}.  
Their measurement of the relative phase of 
the doubly Cabibbo-suppressed $K^{*+}\pi^-$ contribution 
provides encouragement that this may, in future, be a feasible method to 
measure $\phi_3$.

\section{\boldmath{$\sin \left( 2\phi_1 +\phi_3 \right)$} from \boldmath{$B \to D X_u$}}

\subsection{Phenomenology}

The discussion up to this point has centered around $B \to D K$ decays.
However, in each case it is possible to replace the primary kaon with a pion,
and the formalism remains unchanged.
In general, the effect is that the favoured amplitude becomes more favoured, 
and the suppressed amplitude becomes more suppressed.
Consequently the overall rates increase, 
but the sizes of the $CP$ violating effects decrease.
Hence these methods are generally not effective to measure $\phi_3$
(see, however,~\cite{kimoh}).
However, decays of the type $\bar{B}^0 \to D^{(*)+} \pi^{(*)-}$,
where $\pi^* = \rho, a_1$, 
deserve attention, as will be shown.
In this case, the favoured ($b \to c$, {\it eg.} $\bar{B}^0 \to D^{*+} \pi^-$)
and suppressed ($b \to u$, {\it eg.} $\bar{B}^0 \to D^{*-} \pi^+$) 
amplitudes contribute to different final states.
However, since these are neutral $B$ decays, 
there are also contributions from $B^0-\bar{B}^0$ mixing,
and the interference between the suppressed amplitude and the mixing amplitude
leads to $CP$ violating observables~\cite{dstarpi_theory}.

Assuming $CPT$ conservation 
and negligible neutral $B$ meson lifetime difference ($\Delta \Gamma = 0$), 
the generic time-dependent decay rate for a $B$ meson, 
which is tagged as $B^0$ at time $\Delta t = 0$,
to a final state $f$ is given by~\cite{abe}
\begin{eqnarray}
  \lefteqn{
    \Gamma_{B^0 \to f} (\Delta t) = 
    \frac{1}{2} \left| a \right|^2 e^{-\left|\Delta t \right|/\tau_{B^0}} \times }
  \\
  && \hspace{-4mm} \left\{
    (1 + \left| \rho \right|^2) +
    (1 - \left| \rho \right|^2) \cos (\Delta m \Delta t) +
    2 \, {\rm Im}(\rho) \sin (\Delta m \Delta t) \right\}, \nonumber
\end{eqnarray}
whilst that for a $B$ meson tagged as $\bar{B}^0$ is given by
\begin{eqnarray}
  \lefteqn{
    \Gamma_{\bar{B}^0 \to f} (\Delta t) = 
    \frac{1}{2} \left| a \right|^2 e^{-\left|\Delta t \right|/\tau_{B^0}} \times }
  \\
  && \hspace{-4mm} \left\{
    (1 + \left| \rho \right|^2) -
    (1 - \left| \rho \right|^2) \cos (\Delta m \Delta t) -
    2 \, {\rm Im}(\rho) \sin (\Delta m \Delta t) \right\}. \nonumber
\end{eqnarray}
Here $a = A\left(B^0 \to f\right)$ whilst 
$\rho = 
\frac{q}{p}\frac{A\left(\bar{B}^0 \to f\right)}{A\left(B^0 \to f\right)}$.
$\tau_{B^0}$ and $\Delta m$ are the lifetime and mixing parameter of the $B^0$.
Similar equations can be written for the 
decay rates to the conjugate state $\bar{f}$.
Taking $f = D^{*-} \pi^+$ as an example of this class of decays,
note only tree diagrams contribute and so 
$A\left(B^0 \to f\right) = A\left(\bar{B}^0 \to \bar{f}\right)$ and
$A\left(B^0 \to \bar{f}\right) = A\left(\bar{B}^0 \to f\right)$.
Asserting $\left| \frac{q}{p} \right| = 1$, 
identifying 
${\rm arg}\left(\frac{q}{p}\right) = -2\phi_1$, 
$\left|
  \frac{A\left(\bar{B}^0 \to f\right)}{A\left(B^0 \to f\right)}
\right| = R_{D^*\pi}$ and 
${\rm arg}
\left(
  \frac{A\left(\bar{B}^0 \to f\right)}{A\left(B^0 \to f\right)}
\right) = - \phi_3 + \delta_{D^*\pi}$,
and neglecting terms of $R_{D*\pi}^2$, leads to
\begin{eqnarray}
  \lefteqn{\Gamma_{B^0 \to D^{*-}\pi^+} =
    \frac{1}{2} \left| a \right|^2 e^{-\left|\Delta t \right|/\tau_{B^0}} \times \label{eq:b0dsmpip}
  } \\
  &&
  \hspace{2mm}
  \left\{ 
    1 + \cos (\Delta m \Delta t) -
    2 R_{D^*\pi} \sin (\phi_w - \delta_{D^*\pi}) 
    \sin (\Delta m \Delta t) 
  \right\}, \nonumber 
  \\
  \lefteqn{\Gamma_{B^0 \to D^{*+}\pi^-} =
    \frac{1}{2} \left| a \right|^2 e^{-\left|\Delta t \right|/\tau_{B^0}} \times \label{eq:b0dsppim}
  } \\
  &&
  \hspace{2mm}
  \left\{ 
    1 - \cos (\Delta m \Delta t) -
    2 R_{D^*\pi} \sin (\phi_w + \delta_{D^*\pi}) 
    \sin (\Delta m \Delta t) 
  \right\}, \nonumber 
  \\
  \lefteqn{\Gamma_{\bar{B}^0 \to D^{*-}\pi^+} =
    \frac{1}{2} \left| a \right|^2 e^{-\left|\Delta t \right|/\tau_{B^0}} \times \label{eq:b0bdsmpip}
  } 
  \\
  &&
  \hspace{2mm}
  \left\{ 
    1 - \cos (\Delta m \Delta t) +
    2 R_{D^*\pi} \sin (\phi_w - \delta_{D^*\pi}) 
    \sin (\Delta m \Delta t) 
  \right\}, \nonumber 
  \\
  \lefteqn{\Gamma_{\bar{B}^0 \to D^{*+}\pi^-} =
    \frac{1}{2} \left| a \right|^2 e^{-\left|\Delta t \right|/\tau_{B^0}} \times \label{eq:b0bdsppim}
  } \\
  &&
  \hspace{2mm}
  \left\{ 
    1 + \cos (\Delta m \Delta t) +
    2 R_{D^*\pi} \sin (\phi_w + \delta_{D^*\pi}) 
    \sin (\Delta m \Delta t) 
  \right\}, \nonumber
\end{eqnarray}
where the substitution $\phi_w = 2\phi_1 + \phi_3$ has been made.
Therefore, a time-dependent analysis of $B \to D^*\pi$ can yield 
measurements of 
$R_{D^*\pi} \sin (\phi_w + \delta_{D^*\pi})$ and
$R_{D^*\pi} \sin (\phi_w - \delta_{D^*\pi})$.
External information about the value of $R_{D^*\pi}$ is required 
in order to extract $\phi_w$.

Note that the presence of the strong phase $\delta_{D^*\pi}$
hinders the measurement of $\sin(2\phi_1+\phi_3)$, 
adding a fourfold ambiguity.
Ideally, a precise theoretical value for this quantity is desired.
Whilst this may be wishful thinking, 
there are some theoretical arguments that 
strong phases should be small in $B \to D$ decays~\cite{bbns}.

It is often said that since $\sin (2\phi_1)$ is measured 
with good accuracy~\cite{tagging}, a measurement of $\sin(2\phi_1+\phi_3)$
gives the value of $\phi_3$.  
This statement is perhaps misleading, 
since a rather precise measurement of $\sin (2\phi_1)$ may not lead to 
a similarly accurate value of $\phi_3$~\cite{silva03}.
For example, a constraint of $\sin(2\phi_1+\phi_3) > 0.75$
leads approximately to $0^{\circ} < \phi_3 < 90^{\circ}$.
As a corollary, a rather loose constraint,
say $\sin(2\phi_1+\phi_3) < 0.75$, may be able to exclude 
the Standard Model prediction for $\phi_3$.
It is preferable to think of measurements of $\sin(2\phi_1+\phi_3)$
as providing useful constraints on the Unitarity Triangle in their own right.

\subsection{Measurement of \boldmath{$R_{D^*\pi}$}}

In the formalism above,
terms of $R_{D^*\pi}^2$ were neglected,
with the consequence that $R_{D^*\pi}$ could not be measured from the 
time-dependent $D^{*\pm}\pi^{\mp}$ distributions.
Since 
$R_{D^*\pi} 
\approx \left| \frac{V_{cd}^*V_{ub}}{V_{cb}^*V_{ud}} \right| 
\approx 0.02$,
it is clear that the effects of the terms of $R_{D^*\pi}^2$ 
are indeed too small to be observable.
(Note that this is not necessarily the case for a time-dependent analysis
of $D K^0$~\cite{fleischer03}.)
Since there is roughly 20\% theoretical error on the value of 
$R_{D^*\pi}$~\cite{suprun02},
experimental input is desirable.

\begin{figure}
  \hbox to\hsize{
    \hss
    \includegraphics[width=\hsize, bb=25 30 425 359, clip=true]
    {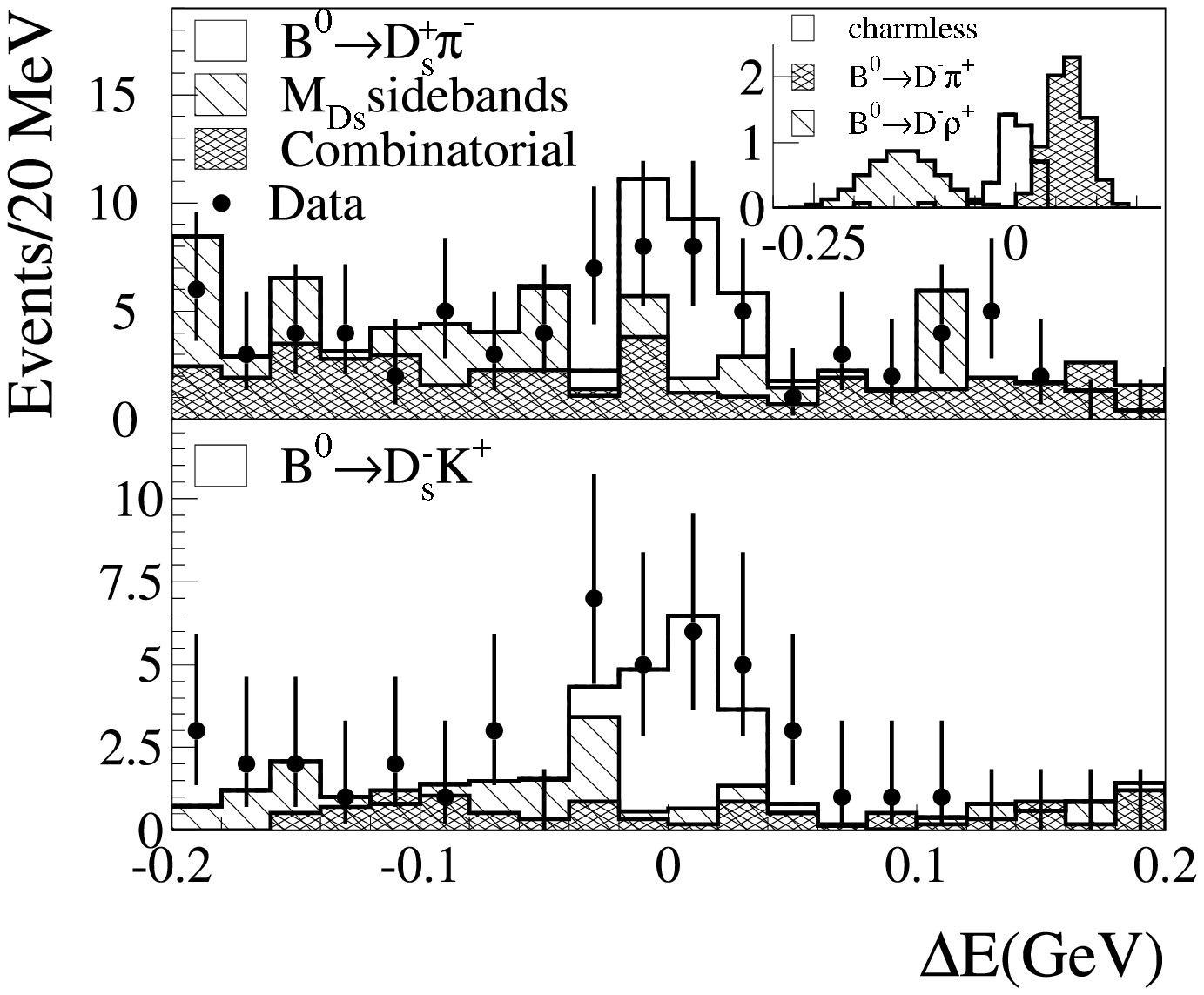}
    \hss}
  \caption{
    Evidence for $\bar{B}^0 \to D_s^- \pi^+$ (top) and 
    $\bar{B}^0 \to D_s^+ K^-$ from BaBar. 
    A data sample of $78\ {\rm fb}^{-1}$ is used.
  }
  \label{fig:babar_dspi}
\end{figure}

\begin{figure}
  \hbox to\hsize{
    \hss
    \includegraphics[width=\hsize]{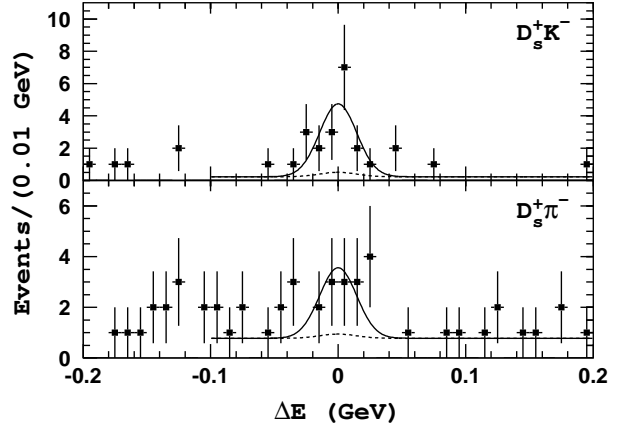}
    \hss}
  \caption{
    Evidence for $\bar{B}^0 \to D_s^+ K^-$ (top) and 
    $\bar{B}^0 \to D_s^- \pi^+$ from Belle. 
    A data sample of $78\ {\rm fb}^{-1}$ is used.
  }
  \label{fig:belle_dspi}
\end{figure}

The decay $B^- \to D^{*-} \pi^0$
is mediated by the suppressed amplitude, and hence 
$\frac
{\Gamma \left( B^- \to D^{*-} \pi^0 \right)}
{\Gamma \left( \bar{B}^0 \to D^{*+} \pi^- \right)} = \frac{1}{2} R_{D^*\pi}^2$
to a good approximation~\cite{dunietz98}.
For the same reason, its branching fraction is extremely small,
and large data samples will be required to measure it
(although the current upper limit of $1.7 \times 10^{-4}$~\cite{pdg2002}
could be dramatically improved with the existing data).
An alternative approach uses the decay $\bar{B}^0 \to D_s^- \pi^+$.
This mode is Cabibbo-enhanced relative to the suppressed amplitude
of interest, and has recently been observed 
as shown in Figs.~\ref{fig:babar_dspi} and~\ref{fig:belle_dspi}.\footnote{
  Evidence for the decay $\bar{B}^0 \to D_s^+ K^-$ is also shown;
  this is only relevant here in that it suggests there may be sizeable
  contributions from $W$-exchange, annihilation or rescattering processes,
  which in turn can affect the extraction of $R_{D^*\pi}$.
}
Since the reconstruction of $\bar{B}^0 \to D_s^- \pi^+$
involves the decay $D_s^- \to \phi \pi^-$,
the poor knowledge of the branching fraction of this decay is responsible
for the limiting systematic error of 
${\mathcal B}\left( \bar{B}^0 \to D_s^- \pi^+ \right)$. 
For this reason, both BaBar and Belle quote the product of branching fractions
${\mathcal B}\left( \bar{B}^0 \to D_s^- \pi^+ \right) \times 
{\mathcal B}\left( D_s^- \to \phi \pi^- \right)$: 
BaBar measure it to be 
$\left(1.13 \pm 0.33 \pm 0.21\right) \times 10^{-6}$~\cite{babar_dspi}
whilst Belle obtain 
$\left(0.86^{+0.37}_{-0.30} \pm 0.11\right) \times 10^{-6}$~\cite{belle_dspi}.
Belle have recently announced a measurement of
${\mathcal B}\left( D_s^- \to \phi \pi^- \right) = 
\left( 3.72 \pm 0.39^{+0.47}_{-0.39} \right) \times 10^{-2}$~\cite{dstophipi},
obtained by comparing the yield of $\bar{B}^0 \to D_s^{*-} D^{*+}$
using a semi-inclusive method ($D_s$ not reconstructed)
to that obtained reconstructing $D_s^-$ in the $\phi \pi^-$ final state.
Combining this value with the World Average~\cite{pdg2002},
averaging the above values for the product branching fraction,
and inserting into the equation
\begin{equation}
{\mathcal B} \left( \bar{B}^0 \to D_s^- \pi^+ \right) \approx
\frac{{\mathcal B} \left( \bar{B}^0 \to D^{*+} \pi^- \right)}{\tan^2 \theta_c}
\left( \frac{f_{D_s}^2}{f_{D^*}^2} \right) R_{D^*\pi}^2
\end{equation}
yields $R_{D^*\pi} \approx 0.022 \pm 0.007$, 
where $f_{D_s}^2/f_{D^*}^2 = 1$ has been assumed.

As will be discussed later, in the vector-vector final state $D^*\rho$
the value of $R_{D^*\rho}$ can in principle be extracted for each 
helicity state.  It may be possible to identify the longitudinally
polarized component as the equivalent value for $D^*\pi$.

\subsection{\boldmath{$B \to D^*\pi$} Time-Dependent Analyses}

\begin{figure}
  \hbox to\hsize{
    \hss
    \includegraphics[width=\hsize]{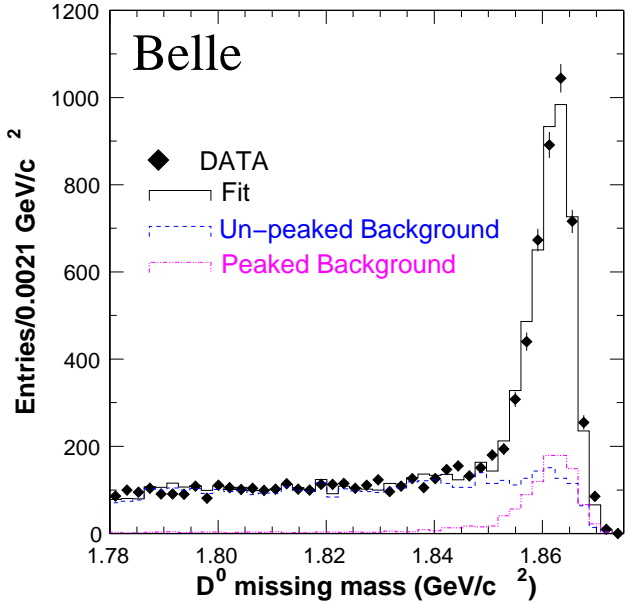}
    \hss}
  \caption{
    Yield of $B \to D^{*}\pi$ events, 
    using a partial reconstruction and lepton tag technique, from Belle.
    Of 4899 candidates, $3433 \pm 81$ are estimated to be signal events.
    A data sample of $30\ {\rm fb}^{-1}$ is used.
  }
  \label{fig:belle_dstarpi}
\end{figure}

\begin{figure}
  \hbox to\hsize{
    \hss
    \includegraphics[width=\hsize, bb=0 345 550 670, clip=true]
    {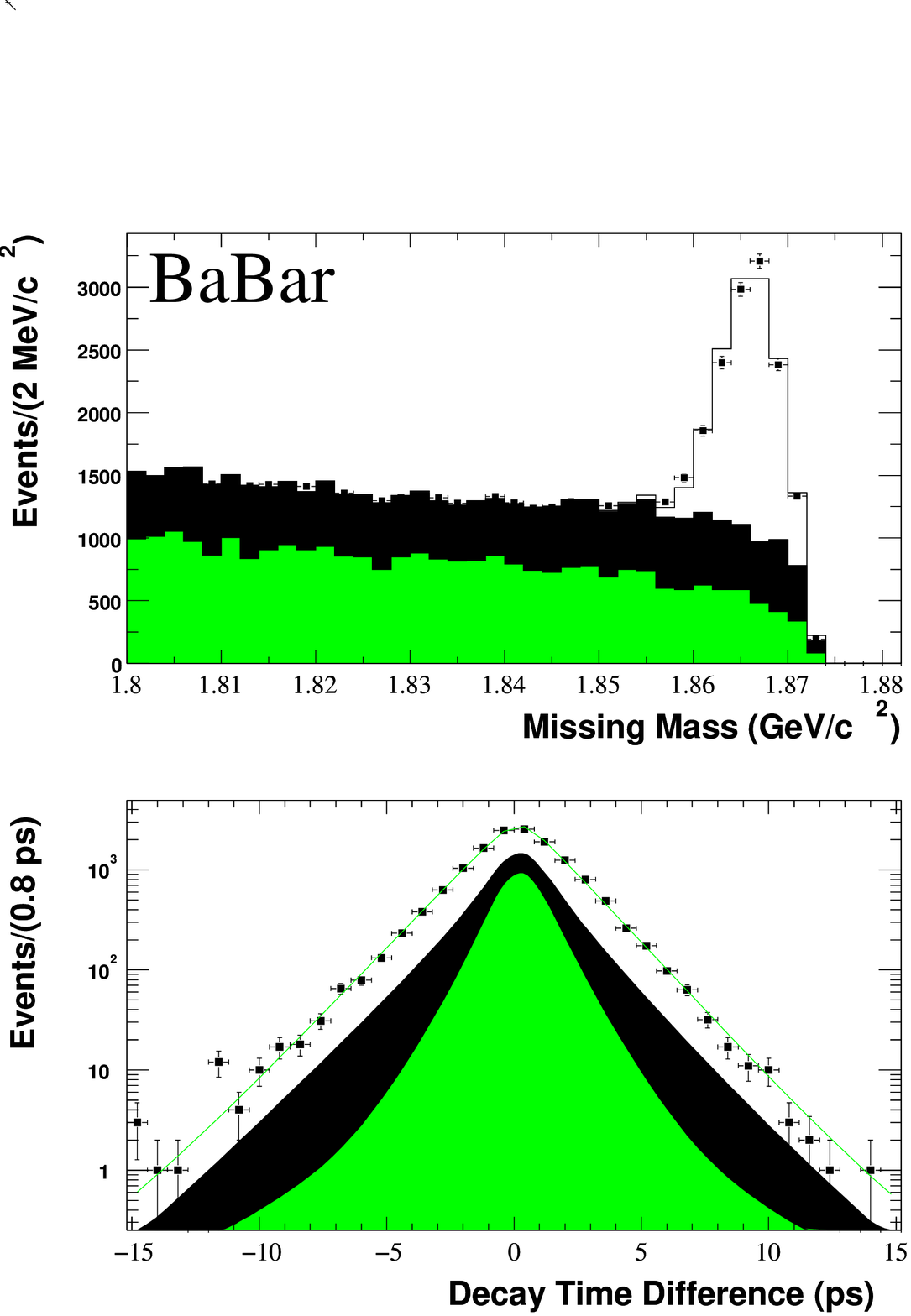}
    \hss}
  \caption{
    Yield of $B \to D^{*}\pi$ events, 
    using a partial reconstruction technique, from BaBar.
    Of approximately 15000 candidates, 
    $6970 \pm 240$ are estimated to be signal events.
    A data sample of $20\ {\rm fb}^{-1}$ is used.
  }
  \label{fig:babar_dstarpi}
\end{figure}

Since the decay $B \to D^*\pi$ 
is amongst the most abundant hadronic $B$ decays,
it has been used in a number of analyses at the $B$ factories.
BaBar, Belle and CLEO have all used this mode to measure
the $B^0$ meson lifetime $\tau_{B^0}$ and the mixing parameter $\Delta m$.
The analysis procedures can be divided into two categories.
The first is full reconstruction, where the $D$ is reconstructed 
in a hadronic final state~\cite{dstarpi_fullrec};
not only $D^*\pi$ but any other similar mode $D^{(*)}\pi^{(*)}$
can, of course, be reconstructed in this manner.
This technique has the advantage of having very little background,
but the efficiency to reconstruct the $D$ is rather low.
To improve the efficiency, a technique called partial reconstruction
can be employed.
Here the decay products of the $D$ are not reconstructed,
but the topology of the prompt (``fast'') pion and that from 
$D^{*+} \to D^0 \pi^+$ decay (``slow'') 
allow separation of signal from background;
clearly this approach can only be used for decay modes 
including a $D^*$ particle.
Analyses which use partial reconstruction tend to suffer from
large backgrounds.  
Similar $B$ meson decays, such as $B \to D^*\rho$, 
can mimic the signal distribution, and are often called ``peaking background''.
Additionally, there is a potentially huge combinatorial background,
which must somehow be controlled.
One approach is to require the presence of a 
high momentum lepton in the event~\cite{cleo_dstarpi_partial}; 
this almost entirely removes background from 
continuum ($e^+e^- \to q\bar{q}, q=u,d,s,c$) processes,
and has the added benefit of cleanly tagging the flavour of the other $B$,
at the cost of sacrificing a large proportion of the signal yield.
Fig.~\ref{fig:belle_dstarpi} shows the $D^*\pi$ candidate events obtained 
by Belle using this technique, 
in an analysis in which the mixing parameter is measured to be 
$\Delta m = 0.509 \pm 0.017 \pm 0.020\ {\rm ps}^{-1}$~\cite{belle_dstarpi_partial}.
Another is to try to separate signal from background from
the topology of the final state particles which are not used
in the reconstruction.
Depending on the precise selection, this approach may
retain a larger signal yield, 
with the price inevitably being larger backgrounds and less clean tags.
Fig.~\ref{fig:babar_dstarpi} shows the $D^*\pi$ candidate events obtained
by BaBar using this technique~\cite{babar_dstarpi_partial},
in an analysis in which the neutral $B$ meson lifetime is measured to be 
$\tau_{B^0} = 1.510 \pm 0.040 \pm 0.041\ {\rm ps}$.

It should be noted that these techniques produce samples
which are approximately independent.
Furthermore, each has different systematic effects.
Therefore, these techniques can be considered as complementary.

From Eqs.~\ref{eq:b0dsmpip}-\ref{eq:b0bdsppim},
it can be seen that non-zero $\sin(2\phi_1+\phi_3)$ 
results in a small $\sin$-like term in the time-evolution of the state.
Rewriting 
$\cos (\Delta m \Delta t) - \epsilon \sin (\Delta m \Delta t) \approx
\cos (\Delta m \Delta t + \epsilon)$,
it is apparent that a small shift in the measured vertex positions,
from which $\Delta t$ is extracted, can mimic $CP$ violation.
At the asymmetric $e^+e^-$ $B$ factories,
a vertex shift of a few microns can have an effect 
of a similar magnitude as that expected of $CP$ violation.
An additional complication is that tagging information is often taken 
from hadronic $B$ decays with the same quark-level process as $D^*\pi$.  
Therefore, these hadronic tags exhibit tag-side $CP$ violation,
which can be as large as that on the signal side~\cite{tagsidecpv}.
In spite of these difficulties, first results on $\sin(2\phi_1+\phi_3)$
are anticipated this summer.

\subsection{\boldmath{$B \to D^* V$} Decays}

Decays of the type $B \to V_1 V_2$, where $V_{1,2}$ represent vector mesons,
have contributions from each of the 
three possible helicity states~\cite{btovv}.
The interference between these amplitudes results in additional 
observables which are sensitive to $CP$ violation~\cite{lss};
in particular in $B \to D^*\rho$ (or $B \to D^* a_1$) decays,
it may be possible to extract $\sin(2\phi_1+\phi_3)$ without
prior knowledge of $R_{D^*\rho}$.
In fact, since there are three helicity states, 
there are three values for $R_{D^*\rho}$ and $\delta_{D^*\rho}$.
Taking into account the relative amplitudes and phases of these contributions, 
there are in total 11 parameters which can, in principle, 
be extracted from the time-dependent angular analysis of these decays.

\begin{figure}
  \hbox to\hsize{
    \hss
    \includegraphics[width=\hsize]{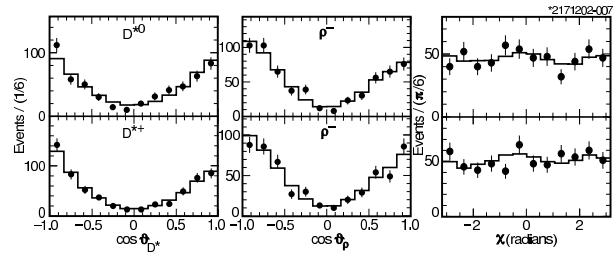}
    \hss}
  \caption{
    Angular analysis of $B \to D^*\rho$ decays by CLEO.
    Left: helicity angle of the $D^*$;
    middle: helicity angle of the $\rho$;
    right: angle between the decay planes of the $D^*$ and the $\rho$.
    Top: $B^- \to D^{*0}\rho^-$; bottom: $\bar{B}^0 \to D^{*+} \rho^-$.
  }
  \label{fig:cleo_dstarrho}
\end{figure}

A first step towards such an analysis is to measure the 
polarization of $B \to D^* \rho$ decays;
this has recently been done by CLEO~\cite{cleo_dstarrho}.
They measure the fraction of longitudinally polarized component
$\Gamma_L/\Gamma = 0.892 \pm 0.018 \pm 0.016$ for $B^- \to D^{*0}\rho^-$ and
$\Gamma_L/\Gamma = 0.885 \pm 0.016 \pm 0.012$ for $\bar{B}^0 \to D^{*+}\rho^-$.
This is illustrated in Fig.~\ref{fig:cleo_dstarrho}.

Whilst time-dependent angular analyses of $B \to VV$ decays,
such as $B \to J/\psi K^{*0}$,
are starting to appear~\cite{belle_jpsikstar},
until now a more popular approach has been to measure the $CP$ content
from the polarization~\cite{babar_dstardstar}.
It may be possible to take a similar approach for
$B \to D^* \rho$ and $B \to D^* a_1$ decays.
This would be extremely beneficial for analyses in which
the final state is partially reconstructed, 
since these techniques tend to result in rather poor angular resolution.
BaBar have used partial reconstruction to reconstruct both 
$D^*\rho$~\cite{babar_dstarpi_partial} and
$D^*a_1$~\cite{babar_dstara1},
obtaining impressive signal yields 
($18400 \pm 1200$ $D^*a_1$ events from a data sample of $20\ {\rm fb}^{-1}$)
but with large backgrounds.
It will be highly challenging to extract $\sin(2\phi_1+\phi_3)$
from these samples.

\section{Summary and Prospects}




At the time of writing, whilst a large number of methods to measure
$\phi_3$ have been proposed, there is no direct experimental input
from $B \to DX$ decays to constrain its value.
$CP$ asymmetries in $B^- \to D_{CP} K^-$ decays have been measured,
but the experimental errors currently cover the entire range of possible
Standard Model values.  
Nevertheless, more experimental information will be forthcoming shortly.
In particular, some of the below may be achieved by summer 2003:
\begin{itemize}
\item Study of the $B^- \to D_{CP}K^-\pi^0$ Dalitz plot
\item Evidence for ADS-style suppressed $D$ decays in
  $B^- \to D K^-$ transitions ({\it eg.} $B^- \to (K^+ \pi^-)_D K^-$)
\item Evidence for $\bar{B}^0 \to \bar{D}^0 \bar{K}^{*0}$
\item Study of multibody $D$ decays in $B^- \to D K^-$ transitions
  ({\it eg.} $B^- \to (K_S \pi^+\pi^-)_D K^-$)
\item Measurement of $\sin(2\phi_1+\phi_3)$ in 
  fully reconstructed $B \to D^{(*)}\pi, D\rho$ decays, 
  and in partially reconstructed $B \to D^*\pi$ decays
\end{itemize}
Therefore one might have the 
first direct experimental constraints on the values of 
$\phi_3$ and $\sin(2\phi_1+\phi_3)$ in such a time scale.

Note however, from a pessimistic viewpoint,
that there is as yet no proof that any of the above methods will succeed!

It may be pragmatic to take a more patient approach and consider what 
will be possible by the time the $e^+e^-$ $B$ factories accumulate
$1000\ {\rm fb}^{-1}$ each.
With such large data sets, each of the methods described above
should be able to provide a useful constraint.
Provided that nature has not been unkind in her choice of strong phases,
direct $CP$ violation in $B$ decays will be established.
Furthermore, the experimental evidence itself will indicate 
which methods are the most promising to precisely measure $\phi_3$ 
and to limit possible ambiguities,
which can also be achieved taking advantage of the redundancy of measurements.
Finally, in such a time scale, 
hadron colliders will be providing measurements of 
$B_{(s)} \to D_{(s)} X$ decays, 
which can be used in a number of ways to measure $\phi_3$.

\end{document}